\documentclass[aps,prl,twocolumn,showpacs]{revtex4-1}
\usepackage{graphicx} 
\usepackage{amssymb}
\usepackage{pifont}
\usepackage{setspace}
\begin{document}
\title
{\bf Exciton Condensates in Double Quantum Wells: A Laboratory for nontrivial topologies and unconventional superconductivity}
\author{Tu\u{g}rul Hakio\u{g}lu$^{\bf (1,2)}$, Mehmet G\"{u}nay$^{\bf (1)}$ and Ege \"{O}zg\"{u}n$^{\bf (1)}$}
\affiliation{${\bf (1)}$ {Department of Physics, Bilkent University, 06800 Ankara, Turkey}
\break
${\bf (2)}$ {Institute of Theoretical and Applied Physics, 48740 Turun\c{c}, Mu\u{g}la, Turkey}}  
\begin{abstract}
It is shown that exciton condensates exhibit an incredible richness in the role that fundamental symmetries play whether manifest or broken. We investigate the appearance of the singlet and the triplet excitonic condensates under spin-orbit coupling (SOC) and consider realistic couplings to the radiation field as manipulated by the fundamental symmetries related to spin, time, orbital and particle degrees of freedom. The role of the strong SOC in inducing a ground state with coexisting singlet and triplet components is shown. It is also shown that the ground state of an excitonic condensate, in the realistic senario of spontaneous coupling to the radiation field, is dominated by a dark component transparent to photoluminescence.    
\end{abstract}
\maketitle
\onecolumngrid
\section{I-Introduction} 
Electron-hole many body interacting systems in semiconductors display a number of unconventional effects at low temperatures. This richness ranges from unconventional forms of superconducting pairing to effects that one would expect to observe in relativistic systems, such as singularities and nontrivial topologies in the energy bands (topological insulators and superconductors). In this article we will concentrate on the condensation of itinerant electron-hole bound states, i.e. the Wannier-Mott excitons that have been speculated a long time ago by Moskalenko, Blatt, Keldish and Kopaev\cite{1,1a,LK1,LK2,LK3} and still is subject to hot debates. A number of experiments have been performed, first in the bulk systems and later in confined geometries, using two-dimensional coupled quantum wells\cite{5,6,9,EC_interactions_general1,EC_interactions_general2,EC_interactions_general3,EC_interactions_general4,10}. The Bose-Einstein Condensation of excitons in a many body interacting environment of itinerant electrons and holes confined in separate planar quantum wells separated by a distance $d$ has been an effective method to understand many of the interesting new phenomena of these systems from both theoretical and experimental perspectives. The need for such confined geometries comes from the fact that the electron-hole ground state formed by the resonant excitation of electrons between a valence and a conduction band of a bulk semiconductor is a metastable bound state with a finite lifetime on the order of a few nanoseconds. Particularly, more recent works in these confined geometries have been carried under zero or strong magnetic fields and, in order to promote the lifetime of the excitons strong external electric fields are also applied by which the exciton lifetime can be increased by a factor of $10^3-10^4$, as compared to their lifetime in the bulk, allowing the thermal equilibrium to be experimentally reached long before they recombine\cite{EC_interactions_general1,EC_interactions_general2,EC_interactions_general3,EC_interactions_general4}. The exciton condensate is a many body collective ground state of electrons and holes, primarily consistent of the fermionic pairs from the s-like electron conduction and the p-like hole valence bands. At low temperatures, the fundamental difference between these systems and the conventional BCS-like pairing\cite{7,7a} is that, the exciton states are formed by nonidentical fermions, i.e. conduction electrons with an effective mass $m_e^*\simeq 6.7\times10^{-2} m_e$ and the p-like heavy holes $m_h^* \simeq 0.4 m_e$ with $m_e$ the bare electron mass. In addition, the p-like bands are composed of light and heavy holes with the light holes being in a higher valence energy band than the heavy holes. These properties have important consequences in the manifestation of the fundamental symmetry operations\cite{8,8a} particularly in the presence of a stabilizing electric field.     

The main experimental tool in search for the exciton condensate has been the photoluminesce technique\cite{5,6,9,10}, and after intense search for years in confined geometries, unquestionable evidence for the condensed state is still lacking \cite{EC_interactions_general1,EC_interactions_general2,EC_interactions_general3,EC_interactions_general4}. An important proposition in this direction was made by Combescot and coworkers\cite{13a} which pointed at the crucial role played by the radiative interband interactions between the electron and the hole bands. Since the Coulomb interaction is spin-neutral, it may be natural to expect that  excitons should come in four degenerate spin configurations. According  to the angular momentum selection rules however, the states (dark excitons) composed of total spin $\pm 2$ do not interact with the radiation field, whereas the states (bright excitons) with spin $\pm 1$ are coupled to it. This residual interaction of the bright pairs to light is not only responsible for their shorter lifetime but it also creates an effective interaction between the electron-hole bands breaking the four-fold degeneracy to two-fold. These radiative processes are dipole-like which can be suppressed in double quantum well (DQW) geometries by a high tunneling barrier between the electron and the hole rich wells. These processes are already well known in the normal electron-hole interacting Fermi liquid phase\cite{Hopfield}, but their investigation in the condensed phase has not been made until recently\cite{TH_EO_SSC}. It is expected that, the residual radiative interaction can be responsible for the inconclusiveness in the observation of the exciton condensate due to the imbalance created in the relative contributions of the dark and the bright components. If the ground state is dominated by the dark excitons, this can bring an explanation to why photoluminescence experiments are not conclusive.   
         
The radiative interband processes can be properly taken into account by including the dipole-field interaction that is present for the bright pairs. The radiation field can then be eliminated using the Markov-Lindblad formalism to obtain the reduced density matrix for the electrons and the holes, yielding an effective interaction of which the unitary part is an effective Hamiltonian for the bright states. We consider the condensate in the self consistent mean field Hartree-Fock scheme in the presence of this effective Hamiltonian.  

From these discussions, we plan to convey the message that, exciton condensation is a broad resource in better understanding some of the unconventional aspects of many body interacting quantum systems from both theoretical and experimental perspectives. In Section.II we outline the many body microscopic model that is used throughout this work and identify manifestation or absence of the basic symmetries in the problem. We also summarize therein, the most commonly worked components of this model (with the radiative transitions and the spin-orbit effects to be discussed in the following sections) and discuss the consequences of the presence or the absence of the relevant fundamental symmetries in the formation of unconventional pairings. Section.III is devoted to the investigation of the effects of the spin-orbit coupling (SOC) and how it enhances pairings with unconventional topologies. In Section.IV, and following Ref.\cite{TH_EO_SSC} a detailed discussion is made on the radiative processes. Using the Markov-Linblad formalism we project the full Hamiltonian onto the fermionic sector which yields an effective reduced Hamiltonian  at the electron-hole level which reflects the spin dependent splitting between the dark and the bright components. The order parameters for the dark and bright states were then solved self consistently using the same method that was employed in Section.I and shown that, the coupling with the radiation field destroys the bright contribution in the condensed ground state.         

\section{II-The Model}
Excitons, unlike a system of interacting pointlike bosons, are composite particles of mutually interacting constituent fermions in a metastable bound state and live in a fluctuating  environment. Unlike point bosons, excitons are affected by the additional mechanism between the constituent fermions. When the wave functions of two excitons overlap, the Pauli exchange mechanism between the constituent fermions in the same quantum well becomes effective which can change the total spins of the excitons between the initial and the final scattering states. Another spin dependent mechanism is the spontaneous coupling to the radiation field which can happen between a conduction electron and a valence hole with opposite spin states. Another important feature is the bulk or the structural inversion asymmetry in many of the semiconductors that are relevant materials for the study of an exciton condensate. Therefore, in the formulation of this problem one has to include these major effects at the microscopic scale. The microscopic model Hamiltonian in this work is given by,      
\begin{eqnarray}
 {\cal H}&=&{\cal H}_0+{\cal H}_{ee}+{\cal H}_{hh}+{\cal H}_{eh}+{\cal H}_{rad}+{\cal H}_{soc}^{(e)}+{\cal H}_{soc}^{(h)}~~~~
\label{hamilt_1}
\end{eqnarray}
where
\begin{eqnarray}
{\cal H}_0&=&\sum_{{\bf k},\sigma}\Bigl\{\xi_{\bf k}^{(e)}\hat{e}_{{\bf k},\sigma}^\dagger \hat{e}_{{\bf k},\sigma}+\xi_{\bf k}^{(h)}\hat{h}_{{\bf k},\sigma}^\dagger \hat{h}_{{\bf k},\sigma}\Bigr\} \\
{\cal H}_{pp}&=&\frac{1}{2}\sum_{{\bf k,k^\prime,q},\sigma,\sigma^\prime}\,{\cal V}_{pp}(q)\hat{p}_{{\bf k}+{\bf q},\sigma}^\dagger \hat{p}_{{\bf k^\prime}-{\bf q},\sigma^\prime}^\dagger \hat{p}_{{\bf k^\prime},\sigma^\prime}\hat{p}_{{\bf k},\sigma} ~~~~~~~~~\\
{\cal H}_{eh}&=&-\sum_{{\bf k,k^\prime,q},\sigma,\sigma^\prime}{\cal V}_{eh}(q)\hat{e}_{{\bf k}+{\bf q},\sigma}^\dagger \hat{h}_{{\bf k^\prime}-{\bf q},\sigma^\prime}^\dagger \hat{h}_{{\bf k^\prime},\sigma^\prime}\hat{e}_{{\bf k},\sigma} \\
{\cal H}_{rad}&=&-\sum_{\bf k} {\bf P}({\bf k}).{\bf E}({\bf k}) \\
{\cal H}_{soc}^{(p)}&=& \sum_{{\bf k},\sigma}\,k\,(e^{i\phi_{\bf k}}\alpha_{p}\hat{p}_{{\bf k},\sigma}^\dagger\hat{p}_{{\bf k},{\bar \sigma}}+h.c.)
\label{hamilt_2}
\end{eqnarray}
with $\sigma, \sigma^\prime$ indicating the individual electron and hole spin states, $p=(e,h)$ indicating the electron (e) or hole (h) degree of freedom, ${\cal H}_0$ is the free Hamiltonian of electrons and holes with $\xi_{\bf k}^{(p)}=\hbar^2k^2/(2m_p)-\mu_p$ indicating the noninteracting single particle energies with band mass $m_p$ measured with respect to individual chemical potentials $\mu_p$, ${\cal H}_{pp}$ includes the complete intraband Coulomb electron-electron (and hole-hole) interaction with ${\cal V}_{pp}(q)=e^2/(2\epsilon q)$, ${\cal H}_{eh}$ is the electron-hole Coulomb interaction with ${\cal V}_{eh}(q)=e^2exp(-q d)/(2\epsilon q)$ where $q=\vert {\bf k}-{\bf k^\prime}\vert)$ and $d$ is the layer separation between the electron and hole quantum wells, ${\cal H}_{rad}$ is the radiative coupling of electrons and holes with dipole strength ${\bf P}({\bf k})$ to the radiation field ${\bf E}({\bf k})$, and finally ${\cal H}_{soc}^{(p)}$ is the spin-orbit coupling for electrons or holes with $k_x+ik_y=ke^{i\phi_{\bf k}}$, and $\alpha_{p}$ as the spin orbit coupling constant. 

The microscopic Hamiltonian in Eq.(\ref{hamilt_1}) is the starting point to solve the full dynamics of the interacting gas of electrons and holes in a DQW. The presence of the radiation field coupling to the excitonic dipoles is a weak correction with nonperturbative consequences. We will not consider these corrections here and handle them in Section.IV. We will also delay discussing the weak SOC terms at the moment. The solution of the sub Hamiltonian ${\cal H}^\prime={\cal H}_0+{\cal H}_{ee}+{\cal H}_{hh}+{\cal H}_{eh}$ in Eq.(\ref{hamilt_1}) was made in Ref.\cite{TH_EO_SSC} using Hartree-Fock mean field (HFMF) approach.  In this approach the Hamiltonian is given by a $4\times 4$ matrix in the electron-hole spinor basis $(\hat{e}_{{\bf k},\uparrow} ~ \hat{e}_{{\bf k},\downarrow} ~ \hat{h}_{-{\bf k},\uparrow}^\dagger ~ \hat{h}_{-{\bf k},\downarrow}^\dagger)$ by 
\begin{eqnarray}
{\cal H}^\prime&=&{\xi}_{\bf k}^{(-)}\,\sigma_0 \otimes \sigma_0+\pmatrix{{\xi}_{\bf k}^{(x)} & {\bf \Delta} ({\bf k}) \cr 
                    {\bf \Delta}^\dagger ({\bf k}) & -{\xi}_{\bf k}^{(x)}\cr}
\label{hamilt_3}
\end{eqnarray}
where $\sigma_0$ is the $2 \times 2$ unit matrix, and in standart notation ${\xi}_{\bf k}^{(-)}=({\xi}_{\bf k}^{(e)}-{\xi}_{\bf k}^{(h)})/2$, ${\xi}_{\bf k}^{(x)}=({\xi}_{\bf k}^{(e)}+{\xi}_{\bf k}^{(h)})/2$. The elements of the Hamiltonian in Eq.(\ref{hamilt_3}) are given by, 
\begin{eqnarray}
\label{hfmf_1}
{\xi}_{\bf k}^{(p)}&=&(\hbar^2{\bf k}^2(2m_p)^{-1}-\mu_p)\sigma_0+{\bf \Sigma}({\bf k})^{(p)} \nonumber \\ 
\Sigma_{\sigma \sigma^\prime}({\bf k})^{(p)}&=&\frac{1}{A}\,\sum_{\bf q}{\cal V}_{pp}({\bf q})\,\langle \hat{p}^\dagger_{{\bf k+q},\sigma}\,\hat{p}_{{\bf k+q},\sigma^\prime}\rangle \,, \qquad {\rm and} \qquad 
\Delta_{\sigma \sigma^\prime}({\bf k})=\frac{1}{A}\,\sum_{\bf q}{\cal V}_{eh}({\bf q})\,\langle \hat{e}^\dagger_{{\bf k+q},\sigma}\,\hat{h}^\dagger_{{\bf -k-q},\sigma^\prime}\rangle 
\end{eqnarray} 
where $\Sigma_{\sigma \sigma^\prime}({\bf k})^{(p)}$ and $\Delta_{\sigma \sigma^\prime}({\bf k})$ are the components of the electron and hole self energies and the order parameter matrix ${\bf \Sigma}({\bf k})^{(p)}$ and ${\bf \Delta}({\bf k})$ respectively. The Eq.'s(\ref{hfmf_1}) are complemented by the electron and hole number conserving conditions yielding $\mu_p$ in Eq.'s(\ref{hfmf_1}). 

Before we discuss this solution, we outline the fundamental symmetries of which the manifestation or the absence play role in the dynamics and simplifies the self consistent solution of Eq.'s(\ref{hfmf_1}). These relevant fundamental symmetries are, the orbital rotation symmetry (OR), space parity (SP), the single and double well spin flip symmetries\cite{SDWSF} (SWSF, DWSF), the time reversal symmetry (TR) and the Fermion Exchange symmetry (FX) which are defined in Table.I. 
\begin{table}[here]
\begin{tabular}{l|l|l|l|l}
\hline
Symm. & $\Sigma_{\sigma \sigma}({\bf k})^{(p)}$ & $\Sigma_{\sigma {\bar \sigma}}({\bf k})^{(p)}$ & $\Delta_{\sigma \sigma}({\bf k})$ & $\Delta_{\sigma {\bar \sigma}}({\bf k})$ \\
\hline TR & $\Sigma^*_{{\bar \sigma} {\bar \sigma}}(-{\bf k})^{(p)}$ & -$\Sigma^*_{{\bar \sigma} \sigma}(-{\bf k})^{(p)}$ & $\Delta_{{\bar \sigma} {\bar \sigma}}^*(-{\bf k})$ & $-\Delta_{{\bar \sigma} \sigma}^*(-{\bf k})$  \\
 SWSF  & $\Sigma_{{\bar \sigma} {\bar \sigma}}({\bf k})^{(p)}$ & -$\Sigma_{{\bar \sigma} \sigma}({\bf k})^{(p)}$ & $\Delta_{{\bar \sigma} \sigma}({\bf k})$ & $\Delta_{{\bar \sigma} {\bar \sigma}}({\bf k})$  \\
 DWSF  & $\Sigma_{{\bar \sigma} {\bar \sigma}}({\bf k})^{(p)}$ & -$\Sigma_{{\bar \sigma} \sigma}({\bf k})^{(p)}$ & $\Delta_{{\bar \sigma} {\bar \sigma}}({\bf k})$ & $-\Delta_{{\bar \sigma} \sigma}({\bf k})$  \\
 SP  & $\Sigma_{\sigma \sigma}(-{\bf k})^{(p)}$ & $\Sigma_{\sigma {\bar \sigma}}(-{\bf k})^{(p)}$ & $\Delta_{\sigma \sigma}(-{\bf k})$ & $\Delta_{\sigma {\bar \sigma}}(-{\bf k})$  \\
 OR  & $\Sigma_{\sigma \sigma}(R:{\bf k})^{(p)}$ & $\Sigma_{\sigma {\bar \sigma}}(R:{\bf k})^{(p)}$ & $\Delta_{\sigma \sigma}(R:{\bf k})$ & $\Delta_{\sigma {\bar \sigma}}(R:{\bf k})$ \\
 FX  & $\Sigma_{\sigma \sigma}({\bf k})^{({\bar p})}$ & $\Sigma_{\sigma {\bar \sigma}}({\bf k})^{({\bar p})}$ & -$\Delta_{\sigma \sigma}(-{\bf k})$ & $-\Delta_{{\bar \sigma} \sigma}(-{\bf k})$  \\
\hline
\end{tabular} 
\caption{Transformations corresponding to the relevant symmetries in this work are depicted. Here ${\bar \uparrow}=\downarrow$ and ${\bar e}=h$ and visa versa.}
\vskip0.3truecm
\end{table}   

In Table.II, the manifestation of the symmetries in Table.I are shown for the microscopic Hamiltonian in Eq.(\ref{hamilt_1}). 
\begin{table}[here]
\begin{tabular}{l|l|l|l|l|l}
\hline
Symm. & ${\cal H}_0$ & ${\cal H}_{pp}$ & ${\cal H}_{eh}$ & ${\cal H}_{rad}$ & ${\cal H}_{soc}^{(p)}$ \\
\hline TR & $\checkmark$ & $\checkmark$ & $\checkmark$ & $\checkmark$ & $\checkmark$ \\
 SWSF  & $\checkmark$ & $\checkmark$ & $\checkmark$ & $\times$ & $\checkmark$   \\
 DWSF  & $\checkmark$ & $\checkmark$ & $\checkmark$ & $\checkmark$ & $\checkmark$   \\
 SP  & $\checkmark$ & $\checkmark$ & $\checkmark$ & $\checkmark$ & $\times$    \\
 OR  & $\checkmark$ & $\checkmark$ & $\checkmark$ & $\checkmark$ & $\times$    \\
 FX  & $\times$ & $\checkmark$ & $\checkmark$ & $\checkmark$ & $\times$ \\
\hline
\end{tabular} 
\caption{The manifestation of the relevant symmetries in Table.I in the microscopic model in Eq.(\ref{hamilt_1}).}
\end{table}
Concentrating on the mean field self-consistent Hamiltonian in Eq.(\ref{hamilt_3}), thus ignoring the radiative corrections and SOC at this moment, it is realized that almost all fundamental symmetries are manifest with the exception of the FX symmetry. In semiconductors there are a number of reasons for the absence of the FX symmetry. In an ordinary semiconductor\cite{topo_exception} the electron-like conduction band is filled by s-like, and the hole-like valence band is filled by the p-like orbitals with intrinsically different band masses. In addition to these fundamental differences between these fermionic bands, the FX symmetry can further be broken extrinsically due to the imbalance in the way that these bands are populated. With all symmetries in Table.II manifest, except the FX symmetry, it is known that the order parameter ${\bf \Delta}({\bf k})$ supports mixed parity pairings\cite{7,7a,8,8a} which is then given by,
\begin{eqnarray}
{\bf \Delta}({\bf k})&=&\pmatrix{-d_x({\bf k})+id_y({\bf k}) & \psi({\bf k})+d_z({\bf k}) \cr 
              -\psi({\bf k})+d_z({\bf k}) & d_x({\bf k})+id_y({\bf k})}=\pmatrix{\Delta_{\uparrow \uparrow}({\bf k}) & \Delta_{\uparrow \downarrow}({\bf k}) \cr 
                      -\Delta_{\uparrow \downarrow}(-{\bf k}) & \Delta_{\uparrow \uparrow}({-\bf k})} 
\label{order_param_2}
\end{eqnarray}
in which, the first matrix is in the Balian-Werthamer notation with ${\bf d}({\bf k})$ denoting the triplet and $\psi({\bf k})$ denoting the singlet spin states\cite{8,8a}. We manifested the TR symmetry in the lower matrix (see Table.I), and in addition, due to the reality of the Coulomb interaction, all matrix elements are real and hence $\psi({\bf k})$ is always even and $d_z({\bf k})$ is always odd under ${\bf k} \to -{\bf k}$. There are further simplifications on the off diagonal matrix elements in Eq.(\ref{order_param_2}). Considering that the model considered has a manifested DWSF symmetry in Table I, we have $\psi({\bf k}) d_z({\bf k}) \propto \vert \Delta_{\uparrow \downarrow}({\bf k})\vert^2 -\vert \Delta_{\downarrow \uparrow}({\bf k})\vert^2=0$. Here the strongly s-wave character of the Coulomb interaction dictates here that $\psi({\bf k})\ne 0$ and $d_z({\bf k})=0$. Note in Table.I that, if the FX symmetry was also manifest, the diagonal components in Eq.(\ref{order_param_2}) would also vanish, leaving only the singlet as in the case of conventional superconductivity.                
  
Finally, the consequence of the manifest TR, SF symmetries and the real interaction is that $\Sigma_{\sigma {\bar \sigma}}({\bf k})^{(p)}=0$ and $\Sigma_{\sigma \sigma}({\bf k})^{(p)}=\Sigma_{{\bar \sigma} {\bar \sigma}}({\bf k})^{(p)}$, hence ${\bf \Sigma}({\bf k})^{(p)}$ is diagonal. Under these simplifications, the eigen energy bands of Eq.(\ref{hamilt_3}) can be easily found as $e_{\bf k}=\xi_{\bf k}^{(-)} \pm E_{\bf k}$ where $E_{\bf k}=\sqrt{({\xi}_{\bf k}^{(x)})^2+{\it Tr}\{{\bf \Delta}{\bf \Delta^\dagger}\}/2}$ with each sign doubly degenerate due to TR symmetry (Kramers degeneracy). A BCS-type generalized ground state of the exciton condensate including unconventional triplet-singlet pairings can also be found analytically. These are not needed in this work, and we will refer the interested reader to Ref.\cite{TH_Robust_GS}.

We are now at the point to complete the self-consistent set Eq.'s (\ref{hamilt_3})and (\ref{hfmf_1}) by the particle number conserving conditions, as given by  
\begin{eqnarray} 
n_x&=&\frac{1}{A}\sum_{\bf k}\,\Bigl\{(1+\frac{\xi_{\bf k}^{(x)})}{E_{\bf k}})(f_+-f_-+1) +(1-\frac{\xi_{\bf k}^{(x)})}{E_{\bf k}})(f_--f_++1)\Bigr\} \\
n_-&=&\frac{1}{A}\sum_{\bf k}\,(f_++f_--1) \nonumber
\label{chempot_1}
\end{eqnarray} 
where $n_x=(n_e+n_h)/2$ and $n_-=(n_e-n_h)/2$ with $n_x$ defined as the exciton concentration and $n_-$ defined as the concentration imbalance in terms of the electron and holes concentrations $n_e$ and $n_h$.  

\section{III-The Spin-Orbit Coupling}
The spin orbit coupling arises from the coupling of the spin degree of freedom $s=1/2$ and the orbital angular momentum $\ell \ne 0$, resulting in a spin dependent splitting $\ell \pm 1/2$ even in the absence of a magnetic field. Unlike the diamond structured pure $Si$ and $Ge$, in inversion asymmetric zinc-blende structures of III-V and II-VI semiconductor compounds, such as $GaAs$, $InSb$ and $Hg_{x}Cd_{1-x}Te$ the center of inversion is intrinsically absent\cite{Winkler}. This inversion asymmetry, often called as the Bulk Inversion Symmetry (BIA), has been known theoretically\cite{BIA_th1,BIA_th2,BIA_th3,BIA_th4} and tested experimentally by analyzing the Shubnikov-de Haas effect\cite{BIA_exp1} as well as the precession of the spin polarization in photoexcited $GaAs$ crystals\cite{BIA_exp2}.  Another type of SOC arises due to structural breaking of the inversion symmetry, often called as the Structural Inversion Symmetry (SIA), arising from the built-in or externally created asymmetries. The presence of the BIA or the SIA is reflected on the breaking of the ${\bf k} \to -{\bf k}$ parity symmetry which is responsible in creating strong charge accumulation and internal crystal electric field in the direction of the inversion breaking. The SIA can also be controlled externally by applying an external electric field which can create strong confining potentials particularly in the position dependent energy band profiles. In both cases the lowest order contribution to the SOC interaction is of first order in ${\bf k}$, and in 2 dimensional QW structure grown in the [001] (z)-direction, it is given in the $(\hat{p}_{{\bf k} \uparrow}~\hat{p}_{{\bf k} \downarrow})$ basis by 
\begin{eqnarray}
{\cal H}_{soc}^{(p)}({\bf k})=\pmatrix{0 & S_{p}({\bf k}) \cr S_{p}^*({\bf k}) & 0\cr}\,,~~~ S_{p}({\bf k})=\alpha_{p}\,k\,e^{i\phi_{\bf k}}~~~~~
\label{H_SOC_1}
\end{eqnarray}
where $p=(e,h)$, $k e^{i\phi_{\bf k}}=(k_x+ik_y)$ and $\alpha_{p}$ is the SOC coupling strength that can be quite different in the electron-like conduction and the hole-like valence bands. In this work, the electron-hole symmetry is strongly broken by the difference of the band masses $m_e$ and $m_h$ and the controllable electron-hole concentration mismatch $n_-=n_e-n_h$. We thus assume for simplicity a real and equal electron-hole symmetric coupling $\alpha_{e}=\alpha_{h}=\alpha=\gamma E_z$ in Eq.(\ref{H_SOC_1}), therefore ${\cal H}_{soc}^{(e)}={\cal H}_{soc}^{(h)}={\cal H}_{soc}({\bf k})$,  where $\gamma$ is a material dependent constant\cite{gamma} and $E_z$ is a strong electric field, either built-in or applied externally, and solve for the energy bands of 
\begin{eqnarray}
{\cal H}^{\prime\prime}={\cal H}^{\prime}+\pmatrix{{\cal H}_{soc}({\bf k}) & 0 \cr 0 & {\cal H}_{soc}^\dagger({\bf k})\cr}
\label{full_H_SOC_1}
\end{eqnarray}  
in the $(\hat{e}_{{\bf k} \uparrow}~\hat{e}_{{\bf k} \downarrow}~\hat{h}_{-{\bf k} \uparrow}^\dagger~\hat{h}_{-{\bf k} \downarrow}^\dagger)$ basis. The self-consistent set of equations corresponding to the exact solution of Eq.'s(\ref{hfmf_1}) and (\ref{chempot_1}) in the ground state of Eq.(\ref{full_H_SOC_1}) can be solved analytically using the manifested symmetries in Table.II. Since SOC preserves the DWSF symmetry as indicated in Table.II, $\psi({\bf k})d_z({\bf k})=0$ is still respected. In this regard, we consider that the presence of SOC does not change the condition $d_z({\bf k})=0$. The energy eigenstates are then given by 
\begin{eqnarray}
e_\lambda^{\pm}({\bf k})={\xi}_{\bf k}^{(-)} \pm E_{\bf k}~,\qquad {\rm where}\qquad E_{\bf k}=\sqrt{({\xi}_{\bf k}^{(x)}\pm \alpha k \gamma_{\bf k} )^2+(\psi_{\bf k}\mp \gamma_{\bf k}F_{\bf k})^2} 
\label{with_soc_e_values_1}
\end{eqnarray}
where $\gamma_{\bf k}=sign(\alpha k {\xi}_{\bf k}^{(x)}-F_{\bf k}\psi_{\bf k})$. In the detailed calculations we found that the triplet expressed as $F_{\bf k}=e^{-i\phi_{\bf k}}\Delta_{\uparrow\uparrow}({\bf k})$, and the singlet $\psi_{\bf k}=\Delta_{\uparrow\downarrow}({\bf k})$ are both real pairing strengths (see Appendix). The exact analytical expressions for the order parameters are given at zero temperature by,
\begin{eqnarray}
\label{with_soc_ord_par_1}
\psi_{\bf k}&=&-\frac{1}{2A} \sum _{\bf q} V_{eh}(q)[u_{11} v_{12}-u^{\ast}_{12} v_{11}+u^{\ast}_{13} v_{14}-u^{\ast}_{14} v_{13}] \nonumber \\
\\
F_{\bf k}&=&-\frac{1}{A} \sum _{\bf q} e^{-i (\phi_{{\bf k}}-\phi_{{\bf k+q}})} V_{eh}(q) [u^{\ast}_{11} v_{11}+u^{\ast}_{13} v_{13}] \nonumber
\end{eqnarray} 
with the coherence factors in the square brackets on the right calculated at momentum ${\bf k+q}$. The derivation of Eq.'s(\ref{with_soc_ord_par_1}) and the isotropic coherence factors are given in the Appendix. The first observation on Eq.'s(\ref{with_soc_ord_par_1}) is that, $\psi_{\bf k}=\psi_k$ and $F_{\bf k}=F_k$ with $k=\vert{\bf k}\vert$ are both isotropic s-wave like and $\Delta_{\uparrow\uparrow}({\bf k})=e^{-i\phi_{\bf k}}\,F_k$ is a p-wave like triplet. Due to the non-zero angular average $\langle exp\{-i (\phi_{{\bf k}}-\phi_{{\bf k+q}})\} V_{eh}(q) \rangle_{\phi_{{\bf k}}-\phi_{{\bf k+q}}}$, the nonlocal interaction $V_{eh}(q)$ allows a nonzero triplet component. The appearance of the triplet component is therefore purely a manifestation of the nonlocal properties of the Coulomb interaction between the electron and the hole quantum wells. 

\begin{figure}[here]

  \centering

  \begin{tabular}{cc}

    \includegraphics[width=80mm]{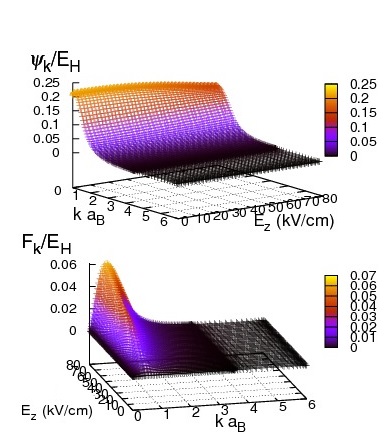}&

    \includegraphics[width=80mm]{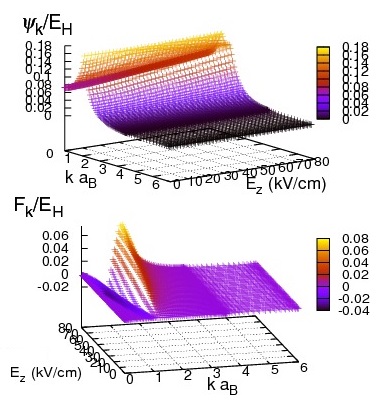}\\
  \end{tabular}
\caption{The singlet, i.e. $\psi_k$, and the triplet, i.e. $F_k$, components of the order parameter, in units of the Hartree energy $E_H\simeq 12 meV$, as a function of $k$ and the electric field $E_z$ (representing here the SOC strength) at zero temperature. The figures in the left column depict the results for low exciton concentration at $n_x a_B^2=0.01$ and the results in the right column are for a relative high concentration at $n_x a_B^2=0.2$. The unit electric field strength $E_z=1 kV/cm$ corresponds in our calculation to $\alpha \simeq 5\times 10^{-5}$ in terms of the $E_H$. The material dependent parameters are adopted for GaAs.}
\label{dark_bright_w_soc}
\end{figure}

Here the triplet solution deserves some attention. A crucial observation here is that, the triplet vanishes in the limit $k \to 0$ linearly, hence $\lim_{k \to 0}\,\Delta_{\uparrow\uparrow}({\bf k})=\beta (k_x+ik_y)$ describes a single vortex. The second observation is that, the proportionality contant $\beta$ decreases linearly with decreasing SOC strength $\alpha$, and in the limit $\alpha \to 0$,  Eq.'s(\ref{with_soc_ord_par_1}) yields a vanishing triplet component. These results are shown in FIG.\ref{dark_bright_w_soc}. On the other hand, in the limit $\alpha \to 0$ one would expect the results of the previous section to be recovered, i.e. $\vert\psi_{\bf k}\vert=\vert \Delta_{\uparrow\uparrow}({\bf k})\vert=F_k$, however we have instead $\vert\psi_{\bf k}\vert\ne 0$ and $\vert \Delta_{\uparrow\uparrow}({\bf k})\vert=F_k=0$. In this limit the eigenvalues are given by $E_{\bf k}=\sqrt{({\xi}_{\bf k}^{(x)})^2+\psi^2_{\bf k}}$ in which case, the energy bands become doubly degenerate but the eigenvectors know about the SOC through their dependence on the phase $\phi_{\bf k}$. This is a manifestation of the fact that, the nonperturbative inversion symmetry breaking is always manifest throughout $\alpha \to 0$, and $\Delta_{\uparrow\uparrow}({\bf k})$ is forced to preserve its single vortex topology all the way down to the limit.  

\break
\section{IV-Radiative Corrections}
Coupling to the radiation field is a fundamental process that cannot be avoided in excitonic systems. In 2D DQWs the effective extend of the quantum wells in the growth (z)-direction can be as large as $W=50 \AA$ when the layer separation $D$ is nearly $a_B\simeq 100 \AA$. The extend and the overlap of the electron and hole wavefunctions along the z-direction is responsible for the finite dipole moment which is the source of the spontaneous dipole-field coupling. The dipole moment is given by 
\begin{eqnarray}
{\bf p}_{0}=\int d{\bf r}\,\Psi_e({\bf r})\,{\bf r} \psi_{h}({\bf r})
\label{dipole_moment}
\end{eqnarray}
described by the overlap integral between the electron $\Psi_e({\bf r})$ and the hole $\Psi_h({\bf r})$ wave functions. In this section we will investigate the effect of this coupling to the radiation field by using the model in Sec.II. One can consider that\cite{TH_EO_SSC}
\begin{eqnarray}
{\bf p}_{0}=e\,d\,exp^{-d^2/W^2} \hat{e}_z 
\label{rad_corr_1}
\end{eqnarray}    
by which the Hamiltonian in Eq.(\ref{hamilt_3}) is replaced by 
\begin{equation} 
{\cal H}^{\prime\prime\prime}={\cal H}^\prime+\sum_{{\bf k},{\bf q},\sigma}\,{\bf p}_{0}\,(\hat{e}_{{\bf k+q},\sigma}^\dagger\hat{h}_{-{\bf k},\bar{\sigma}}^\dagger+h.c.)
\label{rad_corr_2}
\end{equation}
where $\bar{\sigma}$ is the spin opposite to $\sigma$. With the inclusion of ${\cal H}_{rad}$, the new Hamiltonian in Eq.(\ref{rad_corr_2}), due to the anticorrelated electron-hole spins, is no longer invariant under SWSF in Table.II and the effective interaction between the fermions acquires a spin dependent correction as well as a new degree of freedom, i.e. the radiation field. We use the Markov-Lindblad reduced density matrix approach to calculate the effect of the radiation field on the self consistent condensed background. In this formalism the full density matrix $\rho(t)$, composed of the fermionic and radiation field degrees of freedom, is expressed in the interaction picture $\rho^I(t)=exp(i{\cal H}^\prime t)\rho(t)exp(-i{\cal H}^\prime t)$ where ${\cal H}^\prime$ is the exactly solvable part and the ${\cal H}_{rad}$ corresponding to the second term comprises the interaction. The $\rho^I(t)$ respects the equation of motion
\begin{eqnarray}
\label{dens_matr_1}
\frac{\partial \rho^I(t)}{\partial t}=-\frac{i}{\hbar} [{\cal H}_{rad}^I(t),\rho^I(t)]
\end{eqnarray}    
where ${\cal H}_{rad}^I(t)=exp(i{\cal H}^\prime t) {\cal H}_{rad}(t) exp(-i{\cal H}^\prime t)$ is the radiative coupling in the interaction representation. We then project the entire fermionic and field degrees of freedom onto a reduced density matrix ${\rho}_{(R)}^I(t)$ in the fermionic subspace by tracing over the radiation degrees of freedom using the Markov-Lindblad approximation. In this formalism it is possible to find an effective Hamiltonian ${\cal H}^{I_{eff}}_{rad}$ for the fermionic degree o freedom  only, describing the radiative coupling consistently upto the second order in the dipole strength ${\bf p}_0$. In result, Eq.(\ref{dens_matr_1}) can be turned into an effective expression for ${\rho}_{(R)}^I(t)$ as,
\begin{eqnarray}
\frac{\partial {\rho}_{(R)}^I(t)}{\partial t}=-\frac{i}{\hbar} [{\cal H}^{I_{eff}}_{rad}(t),{\rho}_{(R)}^I(t)]
\label{dens_matr_2}
\end{eqnarray}  
the calculation of ${\cal H}^{I_{eff}}_{rad}(t)$ of which the derivation is detailed in Ref.\cite{TH_EO_SSC}. Transforming back to the Schr\"{o}dinger picture, ${\cal H}^{eff}_{rad}$ is given by
\begin{eqnarray}
{\cal H}^{eff}_{rad}\simeq -\sum_{{\bf k},\sigma,\sigma^\prime}\,\Bigl[ g_{\bf k}\,
\langle \hat{e}_{{\bf k},\sigma}^\dagger\,\hat{h}_{{\bf -k},\bar{\sigma}}^\dagger \rangle\,
\hat{h}_{{\bf -k},\bar{\sigma^\prime}} \hat{e}_{{\bf k},\sigma^\prime}+h.c.\Bigr]~~~~~~
\label{eff_hamilt_1}  
\end{eqnarray}
where $g_{\bf k}$ is the effective coupling constant, second order in $p_0=\vert {\bf p}_0\vert$, as given by,  
\begin{eqnarray}
\frac{g_{\bf k}}{g_0}\simeq -\Bigl[1-\frac{1}{2}\frac{\Delta_{\uparrow\uparrow}^2({\bf k})+\Delta_{\uparrow\downarrow}^2({\bf k})}{E_{\bf k}^2}\Bigr]^2~,~ g_0=\frac{2}{\pi}\frac{E_G^2 p_0^2}{\epsilon (\hbar c)^2 d}~~~~~~
\label{eff_coupling_1}  
\end{eqnarray}
where $E_G$ is the energy gap between the electron and the hole bands. Note that ${\cal H}^{eff}_{rad}$ in Eq.(\ref{eff_hamilt_1}) clearly respects the broken SWSF symmetry in Table.II. The energy scale of the radiative coupling, and hence the order of the spin spitting in energy, can be estimated by using the expression for $g_0$ in Eq.(\ref{eff_coupling_1}). For $E_G\simeq 1 eV$, the width of the quantum wells $W\simeq 70 \AA$, the well separation $d\simeq 100 \AA$ a physical energy scale of the radiative corrections can be estimated as $g_0\simeq 10^{-2} meV$.  

We are now at a point where a fermionic effective Hamiltonian can be defined including the effect of the radiative coupling as ${\cal H}^{\prime\prime\prime}_{eff}={\cal H}^{\prime}+{\cal H}^{eff}_{rad}$. In the basis of Eq.(\ref{hamilt_3})
\begin{eqnarray}
{\cal H}^{\prime\prime\prime}_{eff}&=&\pmatrix{{\xi}_{\bf k}^{(x)} & {\bf (\Delta}^{eff})^\dagger ({\bf k}) \cr 
                    {\bf \Delta}^{eff} ({\bf k}) & -{\xi}_{\bf k}^{(x)}\cr}+{\xi}_{\bf k}^{(-)}\,\sigma_0 \otimes \sigma_0~~~~~
\label{eff_total_hamilt}
\end{eqnarray}
where the components of ${\bf \Delta}^{eff}({\bf k})$ are given by,
\begin{eqnarray}
\Delta_{\sigma \sigma^\prime}^{eff}({\bf k})&=&\frac{1}{A}\,\sum_{\bf q}{\cal V}_{eh,\sigma\sigma^\prime}^{eff}({\bf q})\,\langle \hat{e}^\dagger_{{\bf k+q},\sigma}\,\hat{h}^\dagger_{{\bf -k-q},\sigma^\prime}\rangle \nonumber \\
\\
{\cal V}_{eh,\sigma\sigma^\prime}^{eff}({\bf q})&=&{\cal V}_{eh}({\bf q})-\delta_{\sigma^\prime, {\bar \sigma}}\,\delta_{\bf q,0}g_{\bf k} \nonumber
\label{delta_eff_1}
\end{eqnarray} 
here ${\cal V}_{eh,\sigma\sigma^\prime}^{eff}({\bf q})$ is the effective pairing interaction. In addition to the spin-neutral Coulomb contribution, the second term is the effective contribution of the radiative corrections only for anticorrelated electron-hole spins. 

\begin{figure}[here]

  \centering
  
  \begin{tabular}{cc}

    \includegraphics[width=80mm]{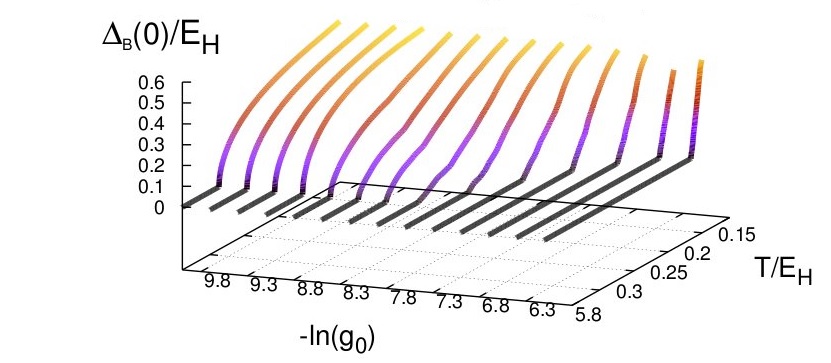}&

    \includegraphics[width=80mm]{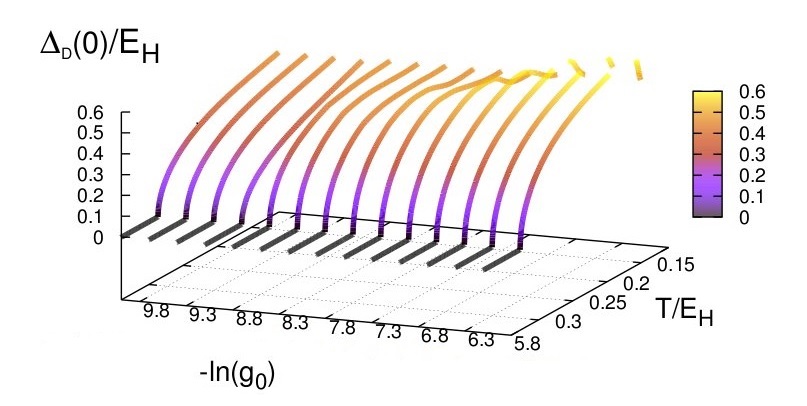}\\
  \end{tabular}
\vskip-0.5truecm
\caption{(Color online) The triplet and the singlet order parameters versus the radiative coupling strength (in log scale) and the temperature adopted from Ref.\cite{TH_EO_SSC}. Here 
$\Delta_D(0)$ and $\Delta_B(0)$ denote $F(k=0)$ and $\psi(k=0)$ in the notation of this work. The solutions look symmetric for vanishingly small $g_0$ which breaks abruptly near $g_0 \simeq 10^{-8} eV \simeq 1.6 \times 10^{-6} E_H$ above which the triplet is nearly stable with a robust critical temperature whereas the singlet is rapidly suppressed. (adopted from Ref.\cite{TH_EO_SSC}).} \vspace{-0.5cm}
\label{fig_delta_eff_1}
\vskip0.5truecm
\end{figure}

Using the same self-consistent scheme in the solution of Eq.'s(\ref{hfmf_1}), and replacing therein $\Delta_{\sigma \sigma^\prime}({\bf k})$ with $\Delta_{\sigma \sigma^\prime}^{eff}({\bf k})$ given by Eq.(\ref{delta_eff_1}), we can solve for the the singlet, i.e. $\Delta_{\uparrow \downarrow}^{eff}({\bf k})$, and the triplet, i.e. $\Delta_{\uparrow \uparrow}^{eff}({\bf k})$ pairing strengths. The results, adopted from Ref.\cite{TH_EO_SSC}, indicate that the singlet contribution is completely destroyed by the presence of the radiative corrections whereas the triplet component survives as shown in FIG.\ref{fig_delta_eff_1}. 

\begin{figure}[here]
\includegraphics[scale=0.42,angle=-90]{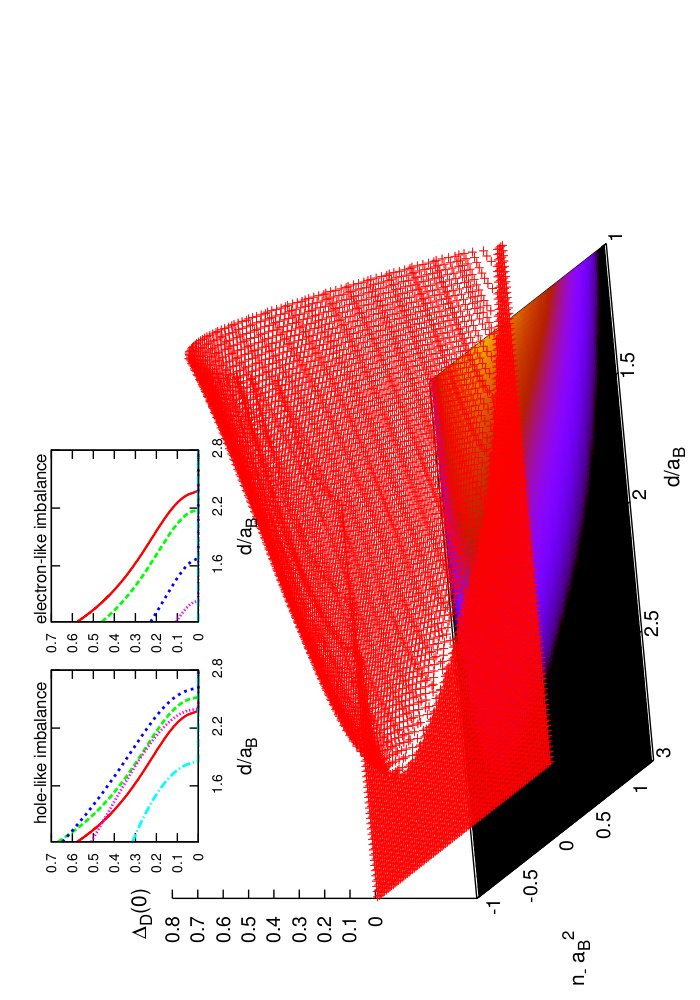}
\caption{(Color online) The triplet order parameter as a function of layer separation $d$ (in units of $a_B^*$) and $n_-$ (in units of $a_B^2$) for $n_x a_B^2=0.8$. The notation is the same as in FIG.\ref{fig_delta_eff_1}. The inlets above are the cross sections of the surface below for hole-like imbalance (left), i.e. $n_- <0$, and the electron-like imbalance (right), i.e. $0 < n_-$. The colors depict: red ($n_-=0$), green ($\vert n_-\vert=0.3$), blue ($\vert n_-\vert=0.8$), purple ($\vert n_-\vert=1.0$), and turquoise ($\vert n_-\vert=1.14$) in units of $1/a_B^2$. (adopted from Ref.\cite{TH_EO_SSC}).} \vspace{-0.5cm}
\label{fig_delta_triplet}
\vskip0.5truecm
\end{figure}

\break
\break
\break
\break
The critical radiative coupling strength $g_0^{(c)}$ was found in Ref.\cite{TH_EO_SSC} to be on the order of $10^{-5} meV$ which implies that, at the physical pairing strength $g_0\simeq 10^{-2} meV$ the singlet component should be absent. The triplet component is shown in FIG.\ref{fig_delta_triplet} as the layer separation $d$ (in units of $a_B$) and the electron-hole concentration imbalance $n_-$ (in units of $1/a_B^2$) are varied. 

\section{VI-Conclusions}
The electron-hole dynamics in semiconductors host many of the most intricate physical processes in many body condensed matter physics. Particularly, in the low temperature regime, the leading processes in understanding the true ground states are not completely resolved. One distinct historical example is the lack of observation of the Wigner crystallization in semiconductors\cite{Monarkha}. On the other hand, from the optimistic side, in semiconductor coupled electron-hole systems one can have a large number of experimentally controllable parameters some of which have been used in this work.  

There are two major conclusions of this work. Firstly, the exact analytical and numerical calculations here indicate that, exciton condensate under strong SOC can have a mixed parity ground state in which the singlet and the triplet components of the condensate coexist with the latter being in a mixed parity state. This is quite contrary to the existing works  in which only singlet or the triplet components are considered. Here we conclude that the ground state is a mix of singlet and a triplet under strong SOC. However, when the spontaneous coupling to the radiation field is correctly taken into account, as our second conclusion, the singlet vanishes leaving the ground state dominantly as a triplet condensate. There may be further implications of a purely triplet condensate. Such as system carries an $h/2$ flux-quantum and can support a nontrivial topology with a Dirac-cone like energy band structure. Thus, yet a new connection can be made with the recently exciting fields of topological insulators\cite{vvv1,vvv2,vvv3} with a controllable topology.         

In this article, we attempted to make a modest case that, exciton condensates pose an opportunity to condensed matter physicist to understand low temperature phenomena ranging from conventional superconductivity to nonconventional spin dependent condensates which can have non trivial topological structures that can even be controlled experimentally in the future. The surprizes are never ending. Recently we realized that the sharp phase boundary depicted in FIG.\ref{fig_delta_triplet} of the dark order parameter as a function of the layer separation $d$ implies that there is an attractive force which, to our knowledge, has not been noticed yet elsewhere. This force, which we coined as the exciton condensate force ${\cal F}_{EC}$, drives its origin from the Coulomb interaction, but its is not the Coulomb force. It reminds the universal Casimir force, but unlike the Casimir force it depends on material parameters and is present only due to the phenomenon of condensation. Our ongoing calculations\cite{EC_force} yield that ${\cal F}_{EC}/A=-3n_x^2/(4\Gamma d_c)$ where $A$ is the sample area, $n_x$ is the exciton concentration, $\Gamma$ is the two dimensional density of free states and $d_c$ is the critical layer separation given by $d_c=e^2\Gamma/(2\epsilon n_x)$. An estimate of this force for a typical $n_x\simeq 10^{11} cm^{-2}$ for GaAs like DQW results in ${\cal F}_{EC}\simeq 10^{-9} N$ for a DQW of a sample area $A\simeq 10^3 \mu m^2$.     

\onecolumngrid

\appendix*
\section{Appendix}
In the 4-component electron-hole basis: $(\hat{e}_{{\bf k} \uparrow} \, \hat{e}_{{\bf k} \downarrow} \, \hat{h}^\dagger_{{\bf -k} \, \uparrow} \hat{h}^\dagger_{{\bf -k} \downarrow})$ the Eq.(\ref{full_H_SOC_1}) is given by,
\begin{eqnarray} 
{\cal H}^{\prime\prime}= \pmatrix{ \hat{A}_{\bf{k}} & \hat{ \Delta }({\bf k}) \cr \hat{ \Delta}^{\dagger} ({\bf k}) & \hat{B}_{\bf{k}}}=\sum_{\bf k}\,\pmatrix{ {\xi}_{\bf k}^{(x)} & S_{e}({\bf k}) & \Delta_{\uparrow \uparrow}({\bf k}) & \Delta_{\uparrow \downarrow}({\bf k}) 
 \cr  S^{\ast}_{e}({\bf k}) & {\xi}_{\bf k}^{(x)} & \Delta_{\downarrow \uparrow}({\bf k}) & \Delta_{\downarrow \downarrow}({\bf k}) 
 \cr \Delta^{\ast}_{\uparrow \uparrow}({\bf k}) & \Delta^{\ast}_{\downarrow \uparrow}({\bf k}) &-{\xi}_{\bf k}^{(x)} & S^{\ast}_{h}({\bf k}) 
 \cr  \Delta^{\ast}_{\uparrow \downarrow}({\bf k}) & \Delta^{\ast}_{\downarrow \downarrow}({\bf k}) & S_{h}({\bf k}) & - {\xi}_{\bf k}^{(x)}} 
\label{Ham1}
\end{eqnarray}
where we ignored the overall constant term in ${\xi}_{\bf k}^{(-)}$ which does not play any role in the diagonalization of Eq.(\ref{full_H_SOC_1}). Since the FX symmetry is broken, we consider the most general unitary transformation for Eq.(\ref{Ham1}), i.e.  $ U{\cal H}^{\prime\prime} = EU$ where $U$ is;

\begin{eqnarray} 
 U=\pmatrix{\hat{u}_{1}(\bf{k}) & \hat{v}_{1}(\bf{k}) \cr \hat{v}_{2}(\bf{k}) & \hat{u}_{2}(\bf{k})}=\pmatrix{u_{11}(\bf{k}) & u_{12}(\bf{k}) & v_{11}(\bf{k})& v_{12}(\bf{k})
 \cr u_{13}(\bf{k}) & u_{14}(\bf{k}) & v_{13}(\bf{k}) & v_{14}(\bf{k})
 \cr v_{21}(\bf{k}) & v_{22}(\bf{k}) & u_{21}(\bf{k}) & u_{22}(\bf{k}) 
 \cr v_{23}(\bf{k}) & v_{24}(\bf{k}) & u_{23}(\bf{k}) & u_{24}(\bf{k})} 
\label{Trans}
\end{eqnarray}
 and $E$ is, 
\begin{eqnarray} 
 E =\pmatrix{E_{e}(\bf{k}) & 0 \cr 0 & E_{h}(\bf{k})} =\pmatrix{E_{e1}(\bf{k}) & 0 & 0 & 0
 \cr 0 & E_{e2}(\bf{k}) & 0 & 0
 \cr 0 & 0 & E_{h1}(\bf{k}) & 0
 \cr 0 & 0 & 0 & E_{h2}(\bf{k}) }
 \end{eqnarray}
Note that ${\it Tr}\{E\}=0$. To find the elements of a $U$-matrix, we have the following set of equations:
\begin{eqnarray} 
\hat{u}_{1} \hat{A} + \hat{v}_{1}  \hat{\Delta}^{\dagger} = E_{e} \hat{u}_{1} \longrightarrow \hat{v}_{1}=(E_{e} \hat{u}_{1}-\hat{u}_{1} \hat{A})[\Delta^{\dagger}]^{-1}  \\ \nonumber
\hat{v}_{1} \hat{B} + \hat{u}_{1}  \hat{\Delta} = E_{e} \hat{v}_{1} \longrightarrow
\hat{u}_{1}  \hat{\Delta}+(E_{e} \hat{u}_{1}-\hat{u}_{1} \hat{A})[\Delta^{\dagger}]^{-1}\hat{B}=E_{e} (E_{e} \hat{u}_{1}-\hat{u}_{1} \hat{A})[\Delta^{\dagger}]^{-1} 
\end{eqnarray}
Defining $ M = [\Delta^{\dagger}]^{-1} \hat{B}  \Delta^{\dagger} =  \pmatrix{P & Q \cr R & T} $ 
\begin{eqnarray} 
\Delta \Delta^{\dagger}=\pmatrix{\vert \Delta_{\uparrow \uparrow}\vert^{2} + \vert \Delta_{\uparrow \downarrow} \vert^{2} & \Delta_{\uparrow \uparrow}  \Delta^{\ast}_{\downarrow \uparrow} +  \Delta_{\uparrow \downarrow}  \Delta^{\ast}_{\downarrow \downarrow} \cr \Delta_{\downarrow \uparrow}  \Delta^{\ast}_{\uparrow \uparrow} +  \Delta_{\downarrow \downarrow}  \Delta^{\ast}_{\uparrow \downarrow} & \vert \Delta_{\downarrow \downarrow}\vert^{2} + \vert \Delta_{\downarrow \uparrow} \vert^{2}}=\pmatrix{n_{1} & n_{2} \cr n_{3}=n^{\ast}_{2} & n_{4}}
\label{Delta}
\end{eqnarray}
we have, 
\begin{eqnarray} 
 P&=&\frac{1}{det(\Delta^{\dagger})} \Big[- {\xi}_{\bf k}^{(x)}  det(\Delta^{\dagger}) -  S_{h} \Delta^{\ast}_{\downarrow \uparrow} \Delta^{\ast}_{\uparrow \uparrow}+ S^{\ast}_{h} \Delta^{\ast}_{\downarrow \downarrow} \Delta^{\ast}_{\uparrow \downarrow}  \Big ]  \\ \nonumber
 Q&=&\frac{-1}{det(\Delta^{\dagger})} \Big[-S^{\ast}_{h} \Delta^{\ast}_{\downarrow \downarrow} \Delta^{\ast}_{\downarrow \downarrow} +S_{h} \Delta^{\ast}_{\downarrow \uparrow} \Delta^{\ast}_{\downarrow \uparrow}\Big ] \\ \nonumber
 R&=&\frac{1}{det(\Delta^{\dagger})} \Big[-S^{\ast}_{h} \Delta^{\ast}_{\uparrow \downarrow} \Delta^{\ast}_{\uparrow \downarrow} +S_{h} \Delta^{\ast}_{\uparrow \uparrow} \Delta^{\ast}_{\uparrow \uparrow} \Big ] \\ \nonumber
 T&=&\frac{1}{det(\Delta^{\dagger})} \Big[- {\xi}_{\bf k}^{(x)}  det(\Delta^{\dagger}) +  S_{h} \Delta^{\ast}_{\downarrow \uparrow} \Delta^{\ast}_{\uparrow \uparrow}- S^{\ast}_{h} \Delta^{\ast}_{\downarrow \downarrow} \Delta^{\ast}_{\uparrow \downarrow} \Big ] \\ \nonumber
 \end{eqnarray}
From the unitarity $U U^{\dagger}=U^{\dagger}U=\sigma_{0} \otimes \sigma_{0}$, we have $u_{12}= \gamma_{1} u_{11} $ and $ u_{13}=\eta_{1} u_{14} $ where $\gamma_{1}=\frac{n_{1}+ (E_{e1}-{\xi}_{\bf k}^{(x)}) (P-E_{e1}) -R S_{e}}{-n^{\ast}_{2}+S^{\ast}_{e} (P-E_{e1})+R({\xi}_{\bf k}^{(x)} -E_{e1})}  $ and $ \eta_{1} = \frac{n_{4}+(E_{e2}-{\xi}_{\bf k}^{(x)} )(T-E_{e2}) -QS^{\ast}_{e} }{-n_{2}+S_{e} (T-E_{e2}) + Q ({\xi}_{\bf k}^{(x)} -E_{e2})}  $, and 
\begin{eqnarray}
\vert u_{11} \vert^{2}&=&\frac{det(\Delta^{\dagger} \Delta)}{\tilde{n}_{1}+\tilde{n}_{2}+\tilde{n}^{\ast}_{2}+\tilde{n}_{4}+det(\Delta^{\dagger} \Delta)(1+\vert \gamma_{1} \vert^{2})}\\
\vert u_{14} \vert^{2}&=&\frac{det(\Delta^{\dagger} \Delta)}{\overline{n}_{1}+\overline{n}_{2}+\overline{n}^{\ast}_{2}+\overline{n}_{4}+det(\Delta^{\dagger} \Delta)(1+\vert \eta_{1} \vert^{2})}\\
\tilde{n}_{1}&=&n_{1}\Big [(\gamma_{1}E_{e1}-\gamma_{1}{\xi}_{\bf k}^{(x)} )(\gamma^{\ast}_{1}E_{e1}-\gamma^{\ast}_{1}{\xi}_{\bf k}^{(x)} -S^{\ast}_{e})\Big ] \\
\tilde{n}_{2}&=&n_{2}\Big [\vert \gamma_{1} \vert^{2} S^{\ast}_{e}(E_{e1}-{\xi}_{\bf k}^{(x)} )+S^{\ast}_{e}(E_{e1}-{\xi}_{\bf k}^{(x)})-\gamma^{\ast}_{1}[(E_{e1}-{\xi}_{\bf k}^{(x)} )^{2}]-\gamma_{1}S^{\ast}_{e}S^{\ast}_{e}     \Big ] \\
\tilde{n}_{4}&=&n_{4}\Big [(E_{e1}-{\xi}_{\bf k}^{(x)} -\gamma^{\ast}_{1}S_{e})(E_{e1}-{\xi}_{\bf k}^{(x)} -\gamma_{1}S^{\ast}_{e})\Big ] \\
\overline{n}_{1}&=&n_{1} \Big [(E_{e2}-{\xi}_{\bf k}^{(x)} - \eta_{1} S_{e})(E_{e2}-{\xi}_{\bf k}^{(x)} - \eta^{\ast}_{1} S^{\ast}_{e}) \Big ]\\
\overline{n}_{2}&=&n_{2}\Big [\vert \eta_{1} \vert^{2} S^{\ast}_{e}(E_{e2}-{\xi}_{\bf k}^{(x)} )+S^{\ast}_{e}(E_{e2}-{\xi}_{\bf k}^{(x)} )-\eta_{1}[(E_{e2}-{\xi}_{\bf k}^{(x)} )^{2}]-\eta^{\ast}_{1}S^{\ast}_{e}S^{\ast}_{e}     \Big ] \\
\overline{n}^{\ast}_{2}&=&n^{\ast}_{2}\Big [ \vert \eta_{1} \vert^{2} S_{e}(E_{e2}-{\xi}_{\bf k}^{(x)} )+S_{e}(E_{e2}-{\xi}_{\bf k}^{(x)} )-\eta^{\ast}_{1}[(E_{e2}-{\xi}_{\bf k}^{(x)} )^{2}]-\eta_{1}S_{e}S_{e}     \Big ] \\
\overline{n}_{4}&=&n_{4} \Big [(\eta^{\ast}_{1}E_{e2}-\eta^{\ast}_{1}{\xi}_{\bf k}^{(x)} -  S_{e})(\eta_{1}E_{e2}-\eta_{1}{\xi}_{\bf k}^{(x)} -  S^{\ast}_{e}) \Big ]
\end{eqnarray}
Defining $ \tilde{M} = [\Delta]^{-1} \hat{A}  \Delta =  \pmatrix{\tilde{P} & \tilde{Q} \cr \tilde{R} & \tilde{T}} $ the lower right elements, i.e. $u_{2i}, (i=1,\dots,4)$ can be found similarly eventually leading to:
\begin{eqnarray} 
 \tilde{P}&=&\frac{1}{det(\Delta)} \Big[ {\xi}_{\bf k}^{(x)}  det(\Delta)+ S_{e} \Delta_{\downarrow \downarrow} \Delta_{\downarrow \uparrow}- S^{\ast}_{e} \Delta_{\uparrow \downarrow} \Delta_{\uparrow \uparrow}  \Big ]  \\ \nonumber
 \tilde{Q}&=&\frac{1}{det(\Delta)} \Big[S_{e} \Delta_{\downarrow \downarrow} \Delta_{\downarrow \downarrow} -S^{\ast}_{e} \Delta_{\uparrow \downarrow} \Delta_{\uparrow \downarrow} \Big ] \\ \nonumber
\tilde{R}&=&\frac{-1}{det(\Delta)} \Big[S_{e} \Delta_{\downarrow \uparrow} \Delta_{\downarrow \uparrow} -S^{\ast}_{e} \Delta_{\uparrow \uparrow} \Delta_{\uparrow \uparrow} \Big ] \\ \nonumber
\tilde{T}&=&\frac{1}{det(\Delta)} \Big[{\xi}_{\bf k}^{(x)}  det(\Delta))+ S^{\ast}_{e} \Delta_{\uparrow \downarrow} \Delta_{\uparrow \uparrow}- S_{e} \Delta_{\downarrow \downarrow} \Delta_{\downarrow \uparrow} \Big ] \\ \nonumber
 \end{eqnarray}
and,
 \begin{eqnarray} 
\Delta^{\dagger} \Delta=\pmatrix{\vert \Delta_{\uparrow \uparrow}\vert^{2} + \vert \Delta_{\downarrow \uparrow} \vert^{2} & \Delta^{\ast}_{\uparrow \uparrow}  \Delta_{\uparrow \downarrow} +  \Delta^{\ast}_{\downarrow \uparrow}  \Delta_{\downarrow \downarrow} \cr \Delta^{\ast}_{\uparrow \downarrow}  \Delta_{\uparrow \uparrow} +  \Delta^{\ast}_{\downarrow \downarrow}  \Delta_{\downarrow \uparrow} & \vert \Delta_{\downarrow \downarrow}\vert^{2} + \vert \Delta_{\uparrow \downarrow} \vert^{2}}=\pmatrix{m_{1} & m_{2} \cr m_{3}=m^{\ast}_{2} & m_{4}}~.
\label{Deltaa}
\end{eqnarray}
All other elements of the $U$ matrix can now be found. Starting with $u_{22}= \gamma_{2} u_{21} $, $ u_{23}=\eta_{2} u_{24} $ where; $ \gamma_{2} = \frac{m_{1}+(E_{h1}+{\xi}_{\bf k}^{(x)})(\tilde{P}-E_{h1}) -\tilde{R} S^{\ast}_{h}}{-m^{\ast}_{2}+S_{h} (\tilde{P}-E_{h1}) + \tilde{R} (-{\xi}_{\bf k}^{(x)} -E_{h1})}  $ and $ \eta_{2} = \frac{m_{4}+(E_{h2}+{\xi}_{\bf k}^{(x)})(\tilde{T}-E_{h2})  -\tilde{Q}S_{h} }{-m_{2}+S^{\ast}_{h} (\tilde{T}-E_{h2}) +\tilde{Q} (-{\xi}_{\bf k}^{(x)} -E_{h2})}  $ we have \\
\begin{eqnarray}
\vert u_{21} \vert^{2}&=&\frac{det(\Delta \Delta^{\dagger})}{\tilde{m}_{1}+\tilde{m}_{2}+\tilde{m}^{\ast}_{2}+\tilde{m}_{4}+det(\Delta \Delta^{\dagger})(1+\vert \gamma_{2} \vert^{2})}\\
\vert u_{24} \vert^{2}&=&\frac{det(\Delta \Delta^{\dagger})}{\overline{m}_{1}+\overline{m}_{2}+\overline{m}^{\ast}_{2}+\overline{m}_{4}+det(\Delta \Delta^{\dagger})(1+ \vert \eta_{2} \vert^{2})}
\end{eqnarray}
and
 \begin{eqnarray}
v_{11}&=&\frac{u_{11}}{det (\Delta^{\dagger})} \Big [ [ \Delta^{\ast}_{\downarrow \downarrow}({\bf k}) - \gamma_{1} \Delta^{\ast}_{\uparrow \downarrow}({\bf k}) ](E_{e1}- {\xi}_{\bf k}^{(x)} )- S^{\ast}_{e}\gamma_{1}\Delta^{\ast}_{\downarrow \downarrow}({\bf k})+S_{e} \Delta^{\ast}_{\uparrow \downarrow}({\bf k}) \Big ]\\
v_{12}&=&\frac{u_{11}}{det(\Delta^{\dagger})} \Big [ [- \Delta^{\ast}_{\downarrow \uparrow}({\bf k}) + \gamma_{1}  \Delta^{\ast}_{\uparrow \uparrow}({\bf k}) ](E_{e1}-{\xi}_{\bf k}^{(x)} ) + S^{\ast}_{e}\gamma_{1}\Delta^{\ast}_{\downarrow \uparrow}({\bf k})-S_{e} \Delta^{\ast}_{\uparrow \uparrow}({\bf k})\Big ]\\
v_{13}&=&\frac{u_{14}}{det(\Delta^{\dagger})} \Big [ [ \Delta^{\ast}_{\downarrow \downarrow}({\bf k})  \eta_{1} - \Delta^{\ast}_{\uparrow \downarrow}({\bf k}) ](E_{e2}- {\xi}_{\bf k}^{(x)} ) - S^{\ast}_{e} \Delta^{\ast}_{\downarrow \downarrow}({\bf k})+S_{e} \eta_{1} \Delta^{\ast}_{\uparrow \downarrow}({\bf k})\Big ]\\
v_{14}&=&\frac{u_{14}}{det(\Delta^{\dagger})} \Big [ [ -\Delta^{\ast}_{\downarrow \uparrow}({\bf k})  \eta_{1} + \Delta^{\ast}_{\uparrow \uparrow}({\bf k}) ](E_{e2}- {\xi}_{\bf k}^{(x)} ) + S^{\ast}_{e} \Delta_{\downarrow \uparrow}({\bf k})-S_{e} \eta_{1} \Delta^{\ast}_{\uparrow \uparrow}({\bf k})\Big ]\\
v_{21}&=&\frac{u_{21}}{det(\Delta)} \Big [ [ \Delta_{\downarrow \downarrow}({\bf k}) - \gamma_{2}  \Delta_{\downarrow \uparrow}({\bf k}) ](E_{h1}+ {\xi}_{\bf k}^{(x)} ) + S^{\ast}_{h} \Delta_{\downarrow \uparrow}({\bf k})-S_{h} \gamma_{2} \Delta_{\downarrow \downarrow}({\bf k})\Big ]\\
v_{22}&=&\frac{u_{21}}{det(\Delta)} \Big [ [- \Delta_{\uparrow \downarrow}({\bf k}) + \gamma_{2}  \Delta_{\uparrow \uparrow}({\bf k}) ](E_{h1}+ {\xi}_{\bf k}^{(x)} ) - S^{\ast}_{h} \Delta_{\uparrow \uparrow}({\bf k})+S_{h} \gamma_{2} \Delta_{\uparrow \downarrow}({\bf k})\Big ]\\
v_{23}&=&\frac{u_{24}}{det(\Delta)} \Big [ [ \Delta_{\downarrow \downarrow}({\bf k}) \eta_{2} - \Delta_{\downarrow \uparrow}({\bf k}) ](E_{h2}+ {\xi}_{\bf k}^{(x)} ) +S^{\ast}_{h} \eta_{2} \Delta_{\downarrow \uparrow}({\bf k})-S_{h}  \Delta_{\downarrow \downarrow}({\bf k}) \Big ]\\
v_{24}&=&\frac{u_{24}}{det(\Delta)} \Big [ [- \Delta_{\uparrow \downarrow}({\bf k}) \eta_{2} + \Delta_{\uparrow \uparrow}({\bf k}) ](E_{h2}+ {\xi}_{\bf k}^{(x)} ) - S^{\ast}_{h} \eta_{2} \Delta_{\uparrow \uparrow}({\bf k})+S_{h}  \Delta_{\uparrow \downarrow}({\bf k}) \Big ]
\end{eqnarray} 
Inserting these relations and the quasiparticle eigen operators in the general definition of $\Delta^{eff}({\bf k})$ in Eq.(\ref{delta_eff_1}), we find the order parameters in Eq.(\ref{with_soc_ord_par_1}).    


\begin{thebibliography}{99}
\bibitem{topo_exception} In topological materials there is a strong band inversion between the s-like and the p-like orbitals in comparison with the most common ordinary semiconductors. This band inversion is due to the strong intrinsic SOC in these materials which are negligible here.
\bibitem{gamma} Typical values for $\gamma$ for different materials such as GaAs, InSb etc. can be deduced from Ref.\cite{Winkler}. In this work, we used $\gamma=5.2 e\AA^2$ for GaAs.
\end{thebibliography}
\end{document}